\documentclass[twocolumn,aps,pre,showpacs]{revtex4}
\usepackage{pslatex}
\usepackage{epsfig}
\usepackage{color}
\usepackage{amsmath}
\newcommand{\gl}[1]{Eq.~(\ref{#1})}

\def\gtrless{\raise2.5pt\hbox{$>$}\llap{\lower2.5pt\hbox{$<$}}}
\def\gtrapprox{\raise2.5pt\hbox{$>$}\llap{\lower2.5pt\hbox{$\approx$}}}
\newcommand{\bsq}[1]{\begin{subequations}\label{#1}}
\newcommand{\esq}{\end{subequations}}
\newcommand{\beq}[1]{\begin{equation}\label{#1}}
\newcommand{\eeq}{\end{equation}}
\newcommand{\beqa}{\begin{eqnarray}}
\newcommand{\eeqa}{\end{eqnarray}}

\renewcommand{\rho}{\varrho}
\def\gtrapprox{\raise2.5pt\hbox{$>$}\llap{\lower2.5pt\hbox{$\approx$}}}
\def\lessapprox{\raise2.5pt\hbox{$<$}\llap{\lower2.5pt\hbox{$\approx$}}}

\begin{document}

\title{Bond formation and slow heterogeneous dynamics in adhesive spheres
with long--ranged repulsion: Quantitative test of Mode Coupling Theory}

\author{O. Henrich$^1$, A.M. Puertas$^2$, M. Sperl$^{1}$,
J. Baschnagel$^3$, and M. Fuchs$^1$}

\affiliation{
$^1$Department of Physics, University of Konstanz,
78457 Konstanz, Germany\\
$^2$Group of Complex Fluids Physics, Department of Applied Physics, University of Almeria,
04120 Almeria, Andaluc\'{\i}a, Spain\\
$^3$Institut Charles Sadron, 6, rue Boussingault, BP 40016, 67083 Strasbourg Cedex, France
}

\date{\today}


\begin{abstract}

A colloidal system of spheres interacting with both a deep and
narrow attractive potential and a shallow long-ranged barrier
exhibits a prepeak in the static structure factor. This peak can be
related to an additional mesoscopic length scale of clusters and/or
voids in the system. Simulation studies of this system have revealed
that it vitrifies upon increasing the attraction into a gel-like
solid at intermediate densities. The dynamics at the mesoscopic
length scale corresponding to the prepeak represents the slowest
mode in the system. Using mode coupling theory with all input
directly taken from simulations, we reveal the mechanism for glassy
arrest in the system at 40\% packing fraction. The effects of the
low-$q$ peak and of polydispersity are considered in detail. We
demonstrate that the local formation of physical bonds is the
process whose slowing down causes arrest.
 It remains largely unaffected by the large-scale
heterogeneities, and sets the clock for the slow cluster mode.
Results from mode-coupling theory without adjustable parameters
agree semi-quantitatively with the local density correlators but
overestimate the lifetime of the mesoscopic structure (voids).

\end{abstract}

\maketitle

\section{Introduction}

Many materials solidify into nonequilibrium structures when particle
interactions become strong and if crystallization proceeds too
slowly. Colloidal dispersions provide a number of model systems to
study these still little understood solidification phenomena as
particle potentials can be tailored and detailed experimental
observations are possible. For example colloidal hard spheres have
been studied intensively, where addition of nonadsorbing polymer
induces an attraction, whose strength and range can be controlled
\cite{Poon2002}. The case of short-ranged attractions has turned out
especially rich. Colloidal particles interacting with already
moderately strong  short-ranged attractions can form metastable
amorphous solids which exhibit exceedingly long life times.

Depending on the colloidal density and attraction strength different types of
metastable arrested solids can be formed. At high densities glasses
are observed when repulsions hinder and even prevent structural
rearrangements \cite{Pusey1986}. Weak short ranged attractions at first melt these
'repulsion-driven glasses' because they distort and loosen the local
packing. At attraction strengths somewhat higher, yet still of the
order of the thermal energy, physical bonds are formed in dense
systems, which leads to aggregation into 'attraction driven
glasses' \cite{Pham2002,Pham2004,Eckert2002}. At attraction strengths high compared to thermal
fluctuations aggregation phenomena proceed far from equilibrium
at low density, resulting in tenuous solids, i.e. gels. At even lower density, the aggregation process leads to cluster formation and aggregation \cite{Segre2001,Sedgwick2004,Lu2006,Gopalakrishnan2007}.

The connection between repulsion and attraction driven glass transitions at high densities
has been understood within a microscopic theoretical framework, namely mode coupling theory (MCT)
\cite{Fabbian1999,Fabbian1999b,Bergenholtz1999,Bergenholtz2000,Dawson2001}. Yet, the mechanisms of
solidification at intermediate attraction strengths and low to
intermediate densities are still not completely understood \cite{Bergenholtz2003,Ahlstrom2007}.
Especially the interplay of equilibrium phase transitions and
aggregation effects is presently under scrutiny \cite{Verduin1995,Shah2003,Manley2005,Sedgwick2004,Lu2006}.
Another aspect that has been considered at low densities is the existence of weak
long-ranged repulsive interactions ('barriers') between the
particles, that may induce density modulations (or non-percolating clusters at low density),
 and thus prevent solidification for some parameter ranges \cite{Stradner2004}.
Small wavevector scattering peaks indicative of mesoscale
correlations have also been observed and employed to characterize
the cluster or gel structures
\cite{Poon1995,Wu2004,Gopalakrishnan2007}. Their observation in
equilibrium systems with long-ranged barriers is expected because
these systems show tendencies to micro-phase separate and to form
interconnected structures of clusters and voids \cite{Sear1999}.
Yet, such long distance correlations and clusters of finite size
were also observed in systems where barriers supposedly are absent
and there presumably have a nonequilibrium origin and arise from
phase separation and/or aggregation and coarsening
\cite{Sciortino1995,Lu2006}. Nonequilibrium origins of the low angle
scattering peaks also were suggested by the observation of their
time evolution \cite{Giglio1992,Poon1995}.  Alternatively, to reach
gelation from the fluid, liquid-gas phase separation can be prevented  setting a
maximum number of bonds per particle \cite{Zaccarelli2006} or total
number of bonds in the system \cite{Hurtado2007}. In these cases,
gelation is directly connected to percolation, and the low-$q$ modes
facilitate the relaxation of the whole system, due to the lack of
stiffness (only two to four neighbours are allowed on average). At
higher density, such modes are expected to be less important than
collective relaxations, and a general theory should account for both
relaxation mechanisms.

In this contribution we test quantitatively the MCT predictions for
low density attractive glasses or high density gels and study the
role of the cluster or void structure on bond formation at an
intermediate density. The input structure factor needed for the MCT
calculations is taken from simulations, and we compare the
properties of the density correlation functions. Additionally, by
comparing the results of MCT calculations for systems with and
without small wavevector pre-peak in the structure, we highlight the
importance of mesoscopic heterogeneities on attraction driven
dynamic arrest. We consider a system of particles interacting with a
narrow attraction and a weak long-ranged repulsion whose dynamics
has been studied intensely by simulations
\cite{Pham2002,Puertas2002,Puertas2003,Puertas2007b}. At the
considered density of 40\% packing fraction, our  system is above
the percolation threshold, and exhibits an equilibrium structural
pre-peak at small wavevectors for parameter ranges where the barrier
suppresses phase-separation. By switching off the repulsive barrier,
the cluster or void structure can be eliminated and the system
becomes homogeneous (below phase separation).

Similar studies, taking the $S_q$ from simulations as input for MCT,
have been performed previously in repulsion driven systems, for
instance  Lennard-Jones \cite{Nauroth1997,Flenner2005}, molecular
glasses \cite{Winkler2000},  silica \cite{Sciortino2001} or sodium
silicate melts \cite{Voigtmann2006}. For attraction driven colloidal
glasses, a similar study has recently been performed by Manley et
al. using experimental structure factors \cite{Manley2005}, and we
reach an analogous conclusion concerning the mechanism of dynamic
arrest. Our study goes beyond Ref.~\cite{Manley2005} because the
input to our MCT calculations is taken from simulations of the
system we address without adjustable parameters, while Manley et al.
adjusted the colloid density  arbitrarily and assumed a specific
form for the static structure factor at large wavevectors. We find
that it is exactly the structure at large wavevectors that dominates
bond formation. Similar effects in the dynamics related to a prepeak
in $S_q$ have also been observed in sodium silicate melts
\cite{Meyer2004,Voigtmann2006}.

\section{Mode Coupling Theory}

This section aims to give a brief overview of the most important results
within MCT concerning the description of liquid-to-glass nonergodicity
transitions \cite{Goetze1991b,Goetze1992}.  We focus on the coherent part
of the density correlation function as it provides most insights into the
physical mechanisms causing glassy arrest.

MCT gives a self-consistent equation of motion of the (normalized)
intermediate scattering function $\Phi_{ q}(t)$, which is the coherent
part of the autocorrelation function of density fluctuations with
wavevector {\bf q} of $N$ particles, defined by
\beq{eq1}
\Phi_{ q}(t) = \frac{1}{N S_q}\sum_{i,j=1}^N \left\langle \exp\{i\,{\bf q}
[{\bf r}_i(t)-{\bf r}_j(0)]\}\right\rangle.
\eeq
The normalization to unity at time $t=0$ is provided by the static
structure factor $S_q=\sum_{i,j=1}^N\langle \exp\{i\,{\bf q} [{\bf
r}_i-{\bf r}_j]\}\rangle/N$, which captures equilibrium density
correlations. The equation of motion for $\Phi_q(t)$ takes the form of a
relaxation equation, where retardation effects with respect to exponential
relaxation on diffusive timescale ${\cal T}_q = \left(D^{\rm s}
q^2/S_q\right)^{-1}$ are contained in a memory kernel $m_q(t)$.
\beqa
{\cal T}_q\dot{\Phi}_q(t)+\Phi_q(t)+\int_0^t dt'
m_q(t-t')\,\dot{\Phi}_q(t')=0\label{eq2}
\eeqa
The initial decay constant ${\cal T}_q$ describes short time diffusive
particle motion, and is set by the short time collective diffusion
coefficient $D^{\rm s}$; it captures instantaneous particle interactions
and will not play an important role at the glass transition.

The central quantity capturing slow structural rearrangements close
to glassy arrest is the memory function $m_q(t)$ which is given in
MCT-approximation:
\beqa m_q(t)={\cal
F}_q[\{\Phi(t)\}]=\frac{1}{2}\sum_{{\bf k p}} V_{\bf q k p}\;
\Phi_k(t)\Phi_p(t)\label{eq3}
\eeqa
The vertices $V_{\bf q k p}$
couple density fluctuations of different wavelengths and thereby
capture a nonlinear feedback mechanism in dense fluids, which is
interpreted as 'cage effect' \cite{Goetze1992}. The memory kernel
can be regarded as a generalized friction kernel, as can easily be
verified after time-Fourier transformation of Eq.~(\ref{eq2}). MCT
is a first principles approach as the vertices are calculated from
the microscopic interactions
\beqa V_{\bf q k p}=S_q S_k S_p
\frac{\varrho^2}{N q^4}[{\bf q\cdot k}\,c_k+{\bf q\cdot p}\, c_p]^2
\delta({\bf q}-{\bf k}-{\bf p})\label{eq4}
\eeqa
The mode coupling
approximation for $m_q$ yields a set of equations that is solved
self-consistently. Hereby the only input to the theory is the static
equilibrium structure factor $S_q$, which enters the memory kernel
$m_q$ directly and via the direct correlation function $c_q
=(1-1/S_q)/\rho$, with $\rho=N/V$ the average density.  According to
MCT the dynamics of the dense fluid is therefore completely
determined by equilibrium quantities plus one scale factor to set the time scale. In the following, we will take
$S_q$ directly from simulations.

Solutions of Eqs.~(\ref{eq2}-\ref{eq4}) show a bifurcation scenario
due to the nonlinear nature of the equations. The bifurcation point
is identified with an idealized liquid-to-glass transition. A
quantity of special interest is the long-time limit of the
normalized density correlator, $f_q=\lim_{t\to\infty}\Phi_q(t)$,
often referred to as  glass form factor or  Edwards-Anderson
nonergodicity parameter. It describes the frozen-in structure of the
glass and obeys
\beq{eq5} \frac{f_q}{1-f_q}={\cal F}_q[f] \,. \eeq
In the fluid regime density fluctuations at different times
decorrelate, so that the long time limit vanishes, $f_q\equiv 0$. On
approaching a critical packing fraction $\phi_{c}$ or a critical
temperature $T_c$, MCT states that strongly coordinated movements
are necessary for structural rearrangements to relax to equilibrium.
MCT identifies two slow structural processes, $\beta$- and
$\alpha$-process, when the glassy structure (described by $f^c_q$)
becomes metastable and takes a long time to relax. In the idealized
picture of MCT, dynamical arrest sets in at the glass transition
with $f_q>0$ when the particles are hindered to escape from their
neighbouring environments.
 This also is accompanied by diverging
relaxation times. Glassy states are  called nonergodic states in
MCT. The value of the glass form factor at the transition is called
critical nonergodicity parameter $f^c_q$.

Although experiments on molecular glass formers have revealed that
the dynamics very close to the transition point is dominated by thermally
activated hopping processes, which the described (idealized) MCT cannot account for, MCT has been
very successful in describing the approach to glassy arrest.  It gives a
quite accurate description of structural relaxation in colloidal dynamics
and there especially of the $\alpha$-process.

 For liquid states and
large times the correlator approaches the $\alpha$-scaling law
$\Phi_q(t)\to\tilde{\Phi}_q(\tilde{t})$ and $\tilde{t}=t/\tau$. Here
the scaling functions $\tilde\Phi_q(\tilde t)$ are independent of
temperature or other control parameters.
 The $\alpha$-relaxation scale $\tau$
is given in MCT by \beq{eq7} \tau = \tau(\epsilon)=
\tau_0\,/|\epsilon|^{\gamma} \mbox{ with }
\gamma=\frac{1}{2a}+\frac{1}{2b} \eeq and depends only on the
separation $\epsilon=\frac{T_c-T}{T_c}$ from the critical point. The
scaling factor $\tau_0$ needs to be determined from matching the
microscopic dynamics and the $\beta$-scaling regime to the
$\alpha$-scaling regime. The anomalous exponents $a$ and $b$ are
solutions of the equation \beq{eq72}
\Gamma^2(1-x)/\Gamma(1-2x)=\lambda \eeq for $0<x=a<1/2$ and $-1\le
x=-b<0$. The exponent parameter $\lambda$ enters in the second order
of the asymptotic expansion of the right-hand side of
Eqs.~(\ref{eq3},\ref{eq5}) around the critical plateau $f_q^c$.
Therefore it depends on the structure factor $S_q^c$ at the critical
point via the vertex Eq.~(\ref{eq4}).

In the vicinity of the critical point, von Schweidler's power-law
describes the initial $\alpha$-relaxation from the plateau to zero.
It is nothing more than the short--time expansion of the
$\alpha$-master functions, which is up to second order \beq{eq8}
\tilde{\Phi}_q(\tilde{t})=f_q^c-h_q\, \tilde{t}^b (1 + k_q
\tilde{t}^b)+{\cal O}(\tilde{t}^{3b}). \eeq The coefficients $h_q$
are called critical amplitude, the $k_q$ are correction amplitudes
\cite{Franosch1997}. Von Schweidler's law is the origin of
stretching (viz.~ non-exponentiality) in the $\alpha$-process of
MCT.

 The final decay
of the structural relaxations in different correlators provides a
definition of the $\alpha$-relaxation times. We use
$\Phi_q(t=\tau_q)=f^c_q/20$. In MCT the increase of the relaxation
times in different correlators is strongly coupled. Their divergence
is inherited directly from the diverging $\alpha$-relaxation scale
$\tau$ of Eq.~(\ref{eq8}). For the $\alpha$-relaxation times of the
density fluctuations in $\Phi_q(t)$ this means a separation into
$\alpha$-scale and a dimensionless factor $\hat{t}_q$ containing the
wavevector dependence. \beq{eq9} \tau_q=\hat{t}_q\,\tau(\epsilon).
\eeq For large wavevectors, $\hat{t}_q\sim q^{-1/b}$ holds in MCT
\cite{Fuchs1994}.

Summarizing this short presentation of MCT, let us note that the
wavevector dependences of the various amplitudes in the asymptotic
MCT predictions will enable us in the following to identify the
physical mechanisms causing glassy arrest. More details about MCT,
the asymptotic expansions and the scaling-laws can be found in
\cite{Goetze1991b,Goetze1992,Franosch1997,Goetze1999}.

\section{Simulation Setup}

Molecular dynamics simulations in the canonical ensemble were
performed considering 1000 quasi-hard particles interacting by a
short range attraction. Because we aim to study the fluid to
non-ergodic transition induced by attractions,  equilibrium phase
transitions, i.e.~crystallization and liquid-gas separation, were
suppressed by suitable choices for the interaction potential, which
we introduce in the following.

The short range attraction mimics the interaction between colloidal
particles induced by non-adsorbing polymers in a colloid polymer
mixture. For monodisperse colloids, this attraction is given by the
Asakura-Oosawa interaction \cite{Poon2002}, which is slightly
modified to include polydispersity \cite{Mendez2000}. The attraction
strength is set by the concentration of polymers, $\phi_p$, and the
range by the polymer size, $\xi$ (see below). This potential has
been slightly corrected near contact $r=d_{12}$ to ensure that the
total interaction potential has the minimum at $d_{12}$
($d_{12}=(a_1+a_2)$, with $a_1$ and $a_2$ the radii of the
interacting particles) \cite{Puertas2003}.
\begin{centering}
\begin{figure}[h]
\centerline{\psfig{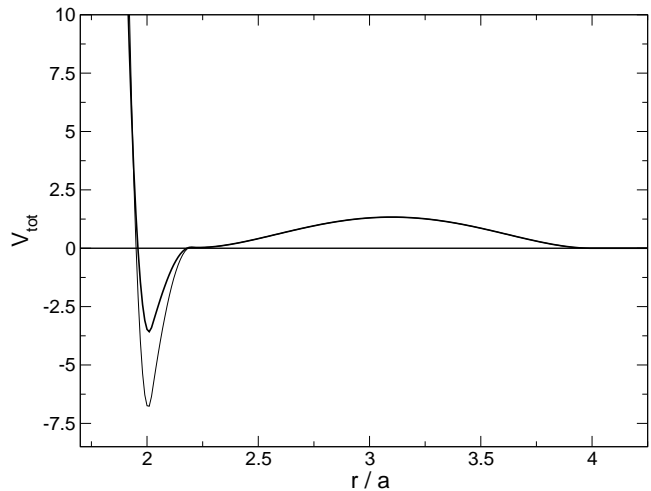}}
\caption{Interaction potential for two particles with the average radius.
The polymer fractions are $\phi_p=0.42$ and $\phi_p=0.25$ for the thin
and thick curves, respectively (values close to the glass transition in
the simulation and MCT, respectively). Note that in our units, the thermal energy is
$k_BT=4/3$.}
\label{potential}
\label{fig1}
\end{figure}
\end{centering}

Crystallization is avoided by using a polydisperse system: particle sizes
are distributed according to a flat distribution of width $\delta=0.1 a$,
where $a$ is the mean diameter. The core-core repulsion is given
by $V_{sc}(r)\:=\:k_BT \left(r/d_{12}\right)^{-36}$.

At high polymer fractions, or attraction strength, this system
separates in two fluid phases with different densities, dilute and
dense -- the critical point is at $\phi_p^{\rm cp}\approx 0.29$. To
avoid this transition, which would interfere with the attractive
glass, a repulsive long-range barrier has been added to the total
interaction potential, which extends from $r=d_{12}+2\xi$ to $r=4a$,
and its height is only $1 k_BT$ (equal to the attraction strength at
$\phi_p=0.0625$). The barrier raises the energy of a dense phase, so
that liquid gas separation does not take place. The resulting total
interaction potential, $V_{tot}=V_{sc}+V_{AO}+V_{bar}$, is
analytical everywhere and allows straightforward integration of the
equations of motions. The total interaction potential is presented
in Fig.  \ref{potential} for particles with the average radius.

The inclusion of the repulsive barrier in the interaction potential
effectively inhibits liquid-gas separation \cite{Puertas2003}, but causes
holes and tunnels in the system. This is reflected in the structure factor
as a low-$q$ peak, which grows and moves to lower-$q$ values, as the
strength of the attraction (namely, $\phi_p$) increases. The effect of
this barrier on the glass transition is studied below within MCT.

Lengths are measured in units of the average radius, $a$, time in units of
$\sqrt{4a^2/3v^2}$, where the thermal velocity $v$ was set to
$\sqrt{4/3}$. Equations of motion were integrated using the
velocity-Verlet algorithm, in the canonical ensemble (constant NTV), to
mimic the colloidal dynamics. Every $n_t$ time steps, the velocity of the
particles was re-scaled to assure constant temperature. No effect of $n_t$
was observed for well equilibrated samples. The time step was set to
$0.0025$. The range of the attraction, $2\xi$, is set to $2\xi=0.2a$. The
density of colloids is reported as volume fraction, $\phi_c=\frac{4}{3}\pi
a^3 \left(1+\left(\frac{\delta}{a}\right)^2\right) n_c$, with $n_c$ the
colloid number density, and the attraction strength is given by the
polymer volume fraction, $\phi_p$ (with $\xi=0.1$, the minimum of the
attraction for average sized particles, at $r=2a$, is $V_{min}=-16 k_BT
\phi_p$).

The dynamics of this system has been analyzed previously within the
framework of MCT, i.e. using the density correlation functions
\cite{Puertas2003,Puertas2005}. Increasing the attraction strength,
$\phi_p$, at $\phi_c=0.40$ a glass transition is obtained at
$\phi_p^c=0.4265$, which shows the qualitative features of
attractive glasses, as predicted by MCT. The critical parameters
given below for the transition were obtained analysing $\Phi_q(t)$
for the fluid state $\phi_p=0.42$, close enough to the glass to show
the typical two-step decay (see Fig. \ref{fig9}) -- 1500 independent
configurations were used to calculate $\Phi_q(t)$, from 15
independent "quenches" from hard spheres at the same density. The
$\beta$-regime and early $\alpha$-decay were analyzed from $t=2$ to
$t=500$. A wavevector-independent von Schweidler exponent was fitted
using the correlators at all wavevectors, whereas the non-ergodicity
parameter, $f_q^c$ and critical amplitudes $h_q$ and $k_q$ were
actually fitted for every wavevector.

The structure factors needed as inputs to MCT were calculated from
the definition of $S_q$, using only the allowed wavevectors ${\bf
q}=2\pi /L (n_x, n_y, n_z)$, with $L$ the box size and $n_x$, $n_y$
and $n_z$ integers. Starting from $q=2\pi/L$, the next value of the
${\bf q}$ modulus is selected if the $q$-separation is larger than
$0.1a^{-1}$, up to $qa=40$. The structure factors were then
interpolated to have a constant $q$-grid.

\section{Aspects of the numerical MCT solutions}

For the numerical solution of the MCT equations,  algorithms
were used that have been developed in the recent years \cite{Fuchs1991b}.

Dynamic and static analyzes were performed by iteratively solving
Eq.~(\ref{eq2})  with the memory functional given by
Eqs.~(\ref{eq3},\ref{eq4}). The results were accepted if a
convergence to $\Phi_q(t)$ with a relative accuracy of $10^{-15}$
was achieved at each $t,q$.  To extend the calculation onto
logarithmic time scales without running into inefficient
time-discretization for late times, we used an algorithm for the
convolution in Eq.~(\ref{eq2}) that doubled the initial time step of
$10^{-9} D^2/D^s$ every 256 time steps, where $D=2a$ is the particle
diameter. The critical polymer concentrations were attained by a
bisection method and determined up to a relative accuracy of
$10^{-5}$ in concentration.

The structure factors for the MCT calculation input were taken
directly from the simulation and linearly interpolated without
further smoothing. We used a wave vector grid with $M=400$ grid
points and a cutoff of $qD=80$. From \cite{Sperl2004} it is known
that with these values neither the discretization nor the cutoff
influence the results significantly. For very small wave vectors
$qD\le0.3$ the algorithm does not produce the correct results
because of numerical error propagation.
 This can be recognized by the static
results shown in Figs. \ref{fig4}-\ref{fig6} for the nonergodicity
parameter $f_q$, in Fig. \ref{fig7} for the critical amplitudes $h_q$ and
in Fig. \ref{fig8} for the $\alpha$-relaxation times. The plotted
results exhibit a sudden drop down to 0 for $q\to0$, whereas they should
take a finite value. Despite this, the results for larger q are not
invalidated, as was verified by removing the incorrect values from the integrations in \gl{eq3}.

\section{Results and Discussion}

In order to quantitatively describe the results from the simulations
for the polydisperse system of particles interacting with the
potential shown in Fig. \ref{fig1}, three different calculations
within MCT were performed. They differ in the input static structure
factors, all of which  were obtained directly from simulations.
Three additional simulation studies were performed solely to
generate the $S_q$ where the pair potentials employed in the
different simulations differed. Thus, we could highlight the
importance of $(i)$ particle polydispersity, and $(ii)$ the
long-ranged repulsive barrier, in the structural relaxation.

For convenience and clarity we name the different MCT calculations
in the following way: System (A) is the {\em monodisperse} model
{\em without repulsive barrier}, whereas we refer with (B) and (C)
to the {\em monodisperse} and {\em polydisperse}  systems,
respectively, {\em with repulsive barrier}. While the calculation in
(C) thus uses exactly the $S_q$ of the system whose dynamics we aim
to describe,  the MCT we use considers a monodisperse system.
Calculation (C) thus also is only approximative. True multi-species MCT calculations like in Ref.~\cite{Foffi2004} would be required to capture all polydispersity aspects, yet are too demanding in the present case. All elements of the matrix of partial structure factors would be required, and the multiple wavevector integrations over the required large $q$-ranges would crucially slow down the MCT numerics.
In order to stress the approximative character of the MCT calculation (C), we call the simulation, where the dynamics is analyzed, system
(D). Some results of the calculations like critical polymer
concentrations, exponent parameters, exponents, and localization
length (in units of the diameter D) are summarized in table
\ref{tab1}.

\begin{centering}
\begin{table}[h]
\begin{tabular}{|*{7}{c|}}
\hline
 & $\delta$ & $V_{bar}$ & $\phi^c_p$ & $\lambda$ & b & $r_l^2$ \\
\hline
\hline
(A) MCT & 0 & No & 0.2356 & 0.752 & 0.555 & 0.00501\\
\hline
(B) MCT & 0 & Yes & 0.2464 & 0.759 & 0.544 & 0.00510\\
\hline
(C) MCT & 0.10 & Yes & 0.3646 & 0.775 & 0.517 & 0.02628\\
\hline
(D) Simulation & 0.10 & Yes & 0.4265 & 0.863 & 0.37 & 0.0158 \\
\hline
\end{tabular}
\caption{Critical polymer concentrations, exponent parameters,
von-Schweidler exponents and localization lengths resulting from the
different MCT computations: (A) monodisperse without barrier, (B)
monodisperse with barrier, (C) polydisperse with barrier. The first
columns give polydispersity $\delta$ and whether a repulsive barrier
exists. Listed under (D) are the corresponding parameters from the
analysis to the polydisperse simulation with barrier of the fluid
state with $\phi_p=0.42$. All states are at colloid packing fraction
$\phi_c=0.40$.} \label{tab1}
\end{table}
\end{centering}

\subsection{Structure factors}

\begin{centering}
\begin{figure}[h]
\centerline{\psfig{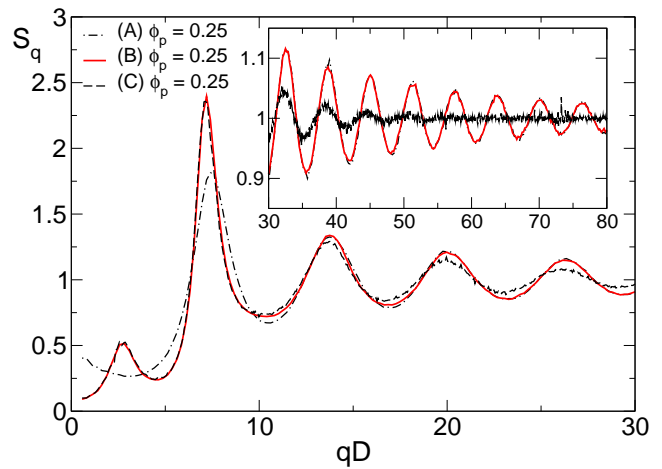}} \caption{$S_q$
from MD-simulations at a colloid packing fraction of $\phi_c=0.40$
and a polymer concentration of $\phi_p=0.25$, which is in the gel
close to the critical point in MCT. The repulsive barrier affects
$S_q$ in the region below $qD\le 13$; system (A) without barrier
(dashed-dotted, black)  shows neither a prepeak nor a primary 
peak which is as high as in systems (B,C) with barrier (full red, dashed black). The
inset demonstrates that polydispersity causes the q-tail
oscillations in $S_q$ to be suppressed for large wavevectors. The
$S_q$ for the polydisperse model (C) falls below the noise level for
$qD\ge45$, whereas the monodisperse systems (A) and (B)  virtually
coincide there.} \label{fig2}
\end{figure}
\end{centering}

\begin{centering}
\begin{figure}[h]
\centerline{\psfig{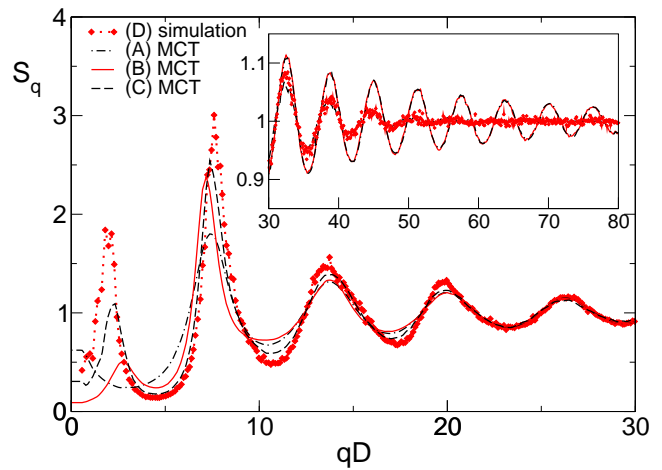}}
\caption{Critical $S_q$ at the boundaries of the gel phase: Black dashed-dotted, red full
and black dashed lines mark results from (A), (B) and (C). The red
diamonds indicate  $S_q$ at $\phi_p=0.42$, which is close to the
arrested state in the simulation (D). In the simulation, the prepeak at $qD\simeq 2.5$ rises and
shifts to smaller $q$-values with increasing polymer concentration. The
inset gives an enlarged view of the q-tail where systems (A,B) show pronounced oscillations driven by the short-ranged attraction.}
\label{fig3}
\end{figure}
\end{centering}

To clarify the influences of the different pair potentials and the
polydispersity on the equilibrium structure we compare on the one
hand structure factors for the same parameters ($\phi_c, \phi_p$)
for the different potentials and on the other hand results at the
MCT-critical points that are obtained for the different potentials.
The $S_q$ in Fig. \ref{fig2} for $\phi_c=0.40$ and $\phi_p=0.25$
show all a primary peak at $qD\approx7.5$ that indicates the local
fluid order. A peculiarity, which is often seen in gels of
intermediate and lower densities, is a low-$q$ peak in the static
structure factor. It appears on the length scale of the voids in the
structure when the sol-gel transition line is reached, increases
towards the transition \cite{Segre2001,Sciortino2004}, and shifts to
slightly lower $q$-values (cf. Fig.  \ref{fig3}). A common
interpretation of the low-$q$ peak is that it indicates the onset of
an arrested phase separation at higher attraction strengths. At
lower density, below the percolation threshold, this peak marks the
presence of clusters in the system, although similar internal
structures for gels and independent clusters have been reported
\cite{Gopalakrishnan2007}. The systems (B) and (C) with repulsive
barrier show this low-$q$ peak, whereas it is absent in the model
(A) without repulsive barrier. The latter $S_q$ clearly grows  in
the limit $q\to0$ for increasing polymer fractions, indicating the
proximity of the liquid-gas  critical point. The $S_q$ with barrier
stay finite as a result of the weak repulsion preventing phase
separation; rather they develop the prepeak which may signal
closeness to microphase separation.

Besides the long-ranged barrier, polydispersity has important
effects on the equilibrium structure. The inset of Fig. \ref{fig2}
displays $S_q$ from $qD=30-80$ for the different systems. The $S_q$
of the monodisperse systems both with and without barrier virtually
coincide and show distinct oscillations for these wave vectors,
unlike in the polydisperse case. Indeed $S_q$ for (C) starts to
deviate from (B) above $qD\approx12$ and decays to the noise level
above $qD\approx45$. This rapid decay to unity in the $S_q$ of the
polydisperse  systems is due to slight differences in the distances
where the (partial) pair correlation functions for differently sized
particles show their contact maxima. This distribution in the
contact distances leads to negative interferences in the oscillatory
large-$q$ pattern in $S_q$, which gets  canceled in the averaged
structure factor of the polydisperse system. This effect of the
short-ranged attraction, viz.~ the increased probability of particle
contact, is thus only contained in the $S_q$ of systems (A) and (B).

Fig.~\ref{fig3} shows $S_q$ at the critical points in MCT (see table
\ref{tab1}) and in the simulation ((D), $\phi^c_p\approx 0.4265$).
It was possible to take the structure factors for the MCT input
directly from the monodisperse simulations, because MCT predicts the
states to be nonergodic already at rather low $\phi_p$, while the
actual system only freezes at higher attraction  strengths. The
features discussed in respect to Fig.~\ref{fig2} can be recognized
again even though the attraction strenghts vary. While $S_q$ of the
system (A) without barrier differs from the other systems at small
wavevectors, polydispersity forces the averaged $S_q$ of systems
(C,D) to approach unity quickly at large wavevectors.

\subsection{Nonergodicity parameters}

From the equilibrium structure factor input of the different systems
we calculated the critical Edwards-Anderson nonergodicity parameters
$f_q^c$ using Eq.~(\ref{eq5}). The bifurcation occurs for the
colloid packing fractions of $\phi_c=0.40$, which has been used
throughout the analysis, at polymer concentrations $\phi^c_p=0.2356$
(A), $\phi^c_p=0.2464$ (B) and $\phi^c_p=0.3646$ (C) respectively.
Note that the critical polymer concentrations are lower than the
critical $\phi^c_p$ in the simulation, system (D). Luckily, the
transition point in system (A) occurs below the critical point,
which suppresses effects from the liquid-gas separation. For system
(B), crystallization is far too slow to affect the results (although
this state is indeed metastable with respect to crystallization).
The trend of MCT to overestimate the tendency to freeze leads almost
to a factor of 2 in terms of attraction strengths \cite{Sperl2004}.
This discrepancy has also been observed in former comparisons of MCT
with binary Lennard-Jones fluids \cite{Nauroth1997}.

\begin{centering}
\begin{figure}[h]
\centerline{\psfig{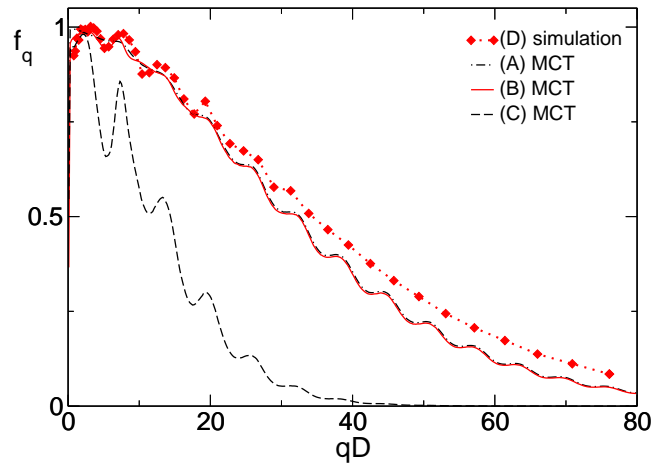}}
\caption{Critical nonergodicity parameters $f^c_q$ at the transition
in the simulation (D) (red
diamonds), from MCT with monodisperse (A,B) $S_q$ (black dashed-dotted and
red full line) and polydisperse (C) $S_q$ (black dashed line). Serious
differences occur, when the average $S_q$ of a polydisperse system
is used as input to a monodisperse theory; the shape of $f_q$ for (C)
resembles more the one of a repulsive glass.}
\label{fig4}
\end{figure}
\end{centering}

The non-ergodicity parameter, $f_q^c$, basically oscillates in phase
with the $S_q$. The shape of $f_q^c$  serves to identify the leading mechanism for the freezing. A repulsion driven transition creates an
$f_q^c$ with pronounced peaks and lower values for small wave vectors. Characteristically, it decreases quickly to zero for increasing $q$. The width of $f_q$ as function of $q$ can be taken as a measure for the localization length that describes the spatial extent that a single particle can explore within its glass cage. For repulsion driven glass transitions one generally finds a localization length of the order of the Lindemann-length, viz.~ a value around a tenth of the average particle separation \cite{Goetze1992}.
If attraction drives the transition, the critical nonergodicity parameters
have higher values and smaller localization lengths, showing up in much
wider $f_q^c$-distributions than for repulsion driven nonergodic states.
The width of $f_q$ as function of $q$ is now set by the attraction range.
In our analysis $f_q^c$ compares quite well with the simulation results
for those models where an attraction driven
 large-q tail is present in  $S_q$ (see Fig.  \ref{fig4}). The
MCT calculation for the polydispersity-smeared-out $S_q$ gives far
too small $f_q^c$ and too large localization lengths what resembles
more results from repulsion driven systems. We conclude from the
agreement between the $f^c_q$  from simulation and the MCT
calculations with attraction-driven large-$q$ tails in $S_q$, that
the simulations exhibit an attraction driven glass transition or
what could be referred to as a gel. Note that this good agreement is
found despite the overestimate of the critical attraction, i.e. the
transition is wrongly located but its principal property is
correctly predicted.

\begin{centering}
\begin{figure}[h]
\centerline{\psfig{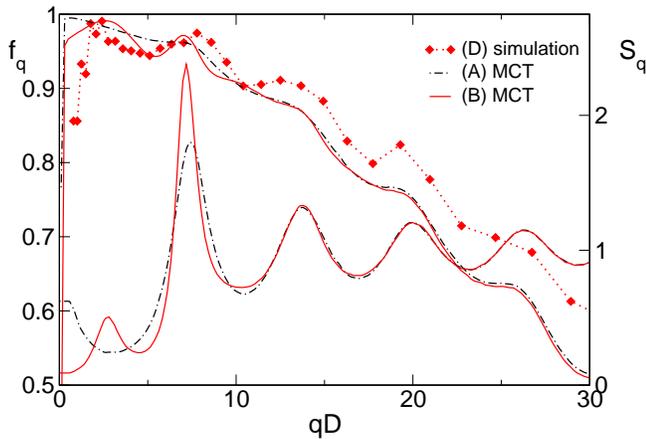}}
\caption{Enlarged view of the nonergodicity parameters of Fig. \ref{fig4} and structure
factors in the low-$q$-region: Minor differences in $f^c_q$
show up in the region $qD\le10$ only, where also the input $S_q$ differs
considerably. Simulation data ((D) red symbols) agree qualitatively with the model calculation (B) including the barrier.}
\label{fig6}
\end{figure}
\end{centering}

An important point to be checked is the role of the long ranged
repulsive barrier. We have already stated above that there is hardly
any difference between the critical polymer concentrations in the
monodisperse calculations without (A) and with barrier (B).
Fig.~\ref{fig6} highlights the influence of the barrier on the
nonergodicity parameters $f_q^c$. The localization length as well as
the attraction driven character of the glass remain unchanged. A
significant effect could only be observed in the vicinity of the
prepeak and the primary peak, where $S_q$ also changes.
Domains of higher wave vectors stay practically unaffected (see also
Fig.~\ref{fig4}). This indicates that the modes on the low angle
q-peak, which is related to the void structure seen in the
simulations, follow the relaxation of an attraction driven glass
without dominating it. It is dominated solely by large-$q$ modes.

\subsection{Critical Amplitudes and Von-Schweidler-Law}

Asymptotic expansions of Eq.~(\ref{eq5}) around the critical plateau
$f_q^c$ introduce the critical amplitude $h_q$ in linear order of
the so-called $\beta$-process. During the $\beta$-process, the
dynamics  on all length scales is strongly coupled. The amplitude
$h_q$ measures the participation of the correlator at wavevector $q$
in this process. The $\beta$-process describes the rate limiting
process of glassy arrest, as here glassy and fluid dynamics start to
deviate. The amplitude $h_q$ thus provides important information on
the physical mechanism causing glassy arrest. It generally exhibits
a minimum as function of $q$ at the maximum of  $f_q^c$, which
describes the stucture that gets frozen in at the glass transition.

From simulation data, $h_q$ can be
 obtained via von-Schweidler fits to correlators. Table \ref{tab1}
reveals that in the present case these fits require somewhat smaller
critical exponents $b$ compared to all MCT calculations (A-C). This
is consistent with the observation, see below in Sect.~E, that the
relaxation appears to be somewhat more stretched in the simulation
than predicted by MCT.
\begin{centering}
\begin{figure}[h]
\centerline{\psfig{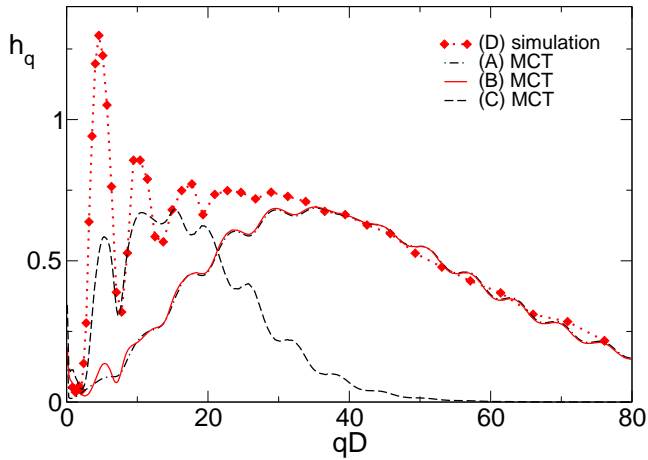}} \caption{Critical
amplitudes in simulation (D) (red diamonds) and MCT (A,B,C) (black
dashed-dotted, red full, black dashed) close to dynamic arrest.
Because of the different $\alpha$-times in simulation and theory the
MCT-results for (A,B) were scaled on the simulation results.}
\label{fig7}
\end{figure}
\end{centering}

The critical amplitudes in Fig.~\ref{fig7}, which are associated
with the calculations where the attractions dominate (A,B), all
exhibit a very broad peak in q. This shows that  very local motion
takes part in the $\beta$-relaxation of an attraction driven glass
transition. The presence (in (B)) or absence (in (A)) of the pre-peak does not influence $h_q$ beyond tiny changes for $qD<10$. MCT thus correctly identifies local bond-formation as
the rate limiting step during the $\beta$-process. MCT
underestimates $h_q$ in the q-range below $qD=30$ and thus
overestimates the stability of the glassy structure on intermediate
and long length scales. Von Schweidler's law, $\Phi_q(t)=f^c_q - h_q
(t/\tau)^b$ from Eq.~(\ref{eq8}), gives a much stronger initial
relaxation of the frozen-in structure for $qD\le30$ in the
simulations than in the MCT calculations.  In the simulation the
small-$q$ modes decay with a larger amplitude during the
$\beta$-process. The overestimate of the stability of the incipient
glassy structure on  length scales larger than corresponding to the
average particle position indicates that MCT misses some of the
larger-distance relaxation mechanisms. Nevertheless, the possibility
to match a common von Schweidler series Eq.~(\ref{eq8}) to the
correlators at large and small wavevectors \cite{Puertas2003}
supports our conclusion from Sect. B that the structural relaxation
for all $q$ is enslaved to local bond formation.

The underestimate of $h_q$ at low $q$ suggests that MCT
underestimates the contribution of the repulsion driven mechanism of
vitrification in the present system. Apparently, in the simulatated
system attraction driven and repulsion driven glass transition
compete, and both transition lines are close. Our quantitative MCT
calculations (erroneously) position the system too far from the
repulsive transitions. Within MCT, a higher order singularity
appears in the vicinity of the merging of the two glass transition
lines, signalled by $\lambda=1$. Indeed, a larger $\lambda$ exponent
is observed in the simulations compared to the MCT calculations,
which glassify due to bond formation solely.

\subsection{$\alpha$-relaxation times}

\begin{centering}
\begin{figure}[h]
\centerline{\psfig{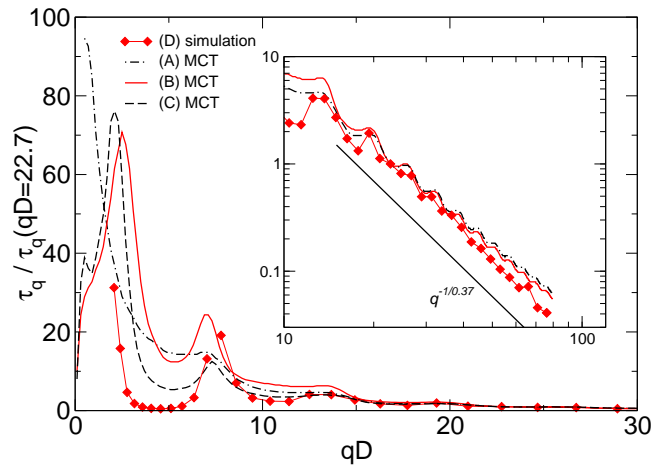}}
\caption{$\alpha$-relaxation times $\tau_q$ in MCT for the systems
(A),(B),(C) (black dashed-dotted, red full, black dashed line) and
in the polydisperse simulation ((D) red diamonds). The curves are
normalized to the value of $\tau_q$ at $qD=22.7$. In the inset, the
results for models (A) and (B) and simulation (D) are shown in a
log-log plot. The straight line gives the asymptotic behavior
$\tau_q=q^{-1/b}$ valid for large $q$ with the corresponding von
Schweidler $b=0.37$ obtained previously.} \label{fig8}
\end{figure}
\end{centering}

The criterion for quantitatively defining the $\alpha$-relaxation
times is somewhat arbitrary. We chose the definition \beq{eq11}
\Phi_q(\tau_q)=X\cdot f_q, \eeq for the $\alpha$-relaxation times,
where in the theory $X=0.05$ and in the simulation $X=0.50$. The latter
choice was required because of the limited simulation time, but
incurs larger corrections to the values of $\tau_q$ arising from
faster relaxation processes. The different definitions are reconciled in the following comparison by normalization of the times, $\tau_q/\tau_{qD=22.7}$, which brings out the $q$-dependence. The $\alpha$-relaxation times $\tau_q$
generally vary in phase with the nonergodicity parameter $f_q^c$ and
the structure factor $S_q$, a phenomenon often referred to as
de~Gennes narrowing. Repulsion driven glass transitions display the
largest $\tau_q$ at the principal peak in $S_q$, which indicates
that the cage formed by the particle's next neighbors induces the
dynamical arrest. Here on the contrary, the slowest relaxation takes
place either (i) at the prepeak, when the barrier causes void
formation,  or (ii) for $q\to0$ in model (A) without barrier on
approaching the phase separation region.

 Figure \ref{fig8}
gives $\alpha$-relaxation times calculated from the above definition
and normalized to their value at $qD=22.7$. In MCT for models with
barrier (B,C) one finds that the
slowest modes are connected with the prepeak. In model (A) $\tau_q$
decreases by more than a factor 2 at this $q$, when the prepeak in
$S_q$ is eliminated. The simulations, however, do not allow to check this difference, but only show that the slowest modes are those with $qD<2$.
Nevertheless, the different dynamics at small $q$ has no further  
impact on the dynamics at larger $q$; in the inset of Fig.~\ref{fig8} the $\tau_q$
for all models agree at large $q$, where the power-law $\tau_q\sim
q^{-1/b}$ is also tested. It holds nicely in the simulation data,
with the von Schweidler exponent obtained from the fitting of the
correlation functions: $b=0.37$. MCT explains $\tau_q$ for larger
wave vectors quite well, though small systematic deviations emerge
because of the difference in the von-Schweidler exponents between
simulation and calculations, models (A), (B) and (C). Let us note in passing that the asymptotic behavior $\tau_q\sim q^{-1/b}$ holds earlier in the polydisperse MCT calculation (C) (not shown), then in the monodisperse ones (A) and (B), where deviations are still noticeable in the inset of Fig.~\ref{fig8}. This appears to support the probabilistic interpretation of the Kohlrausch law within MCT \cite{Goetze1992,Fuchs1994}.

 We conclude
that the void structure is completely enslaved by the bond formation
on local length scales. Even though the dissolution of the void
structure is the slowest process, there's no evidence for a
significant influence of the voids on the local dynamics. Local bond
formation proceeds identically in systems (B,C,D) with barrier and
void pre-peak in $S_q$, and in system (A) without barrier and
void-correlation peak in $S_q$.

\subsection{Correlation functions}

For an inclusive check of the results beyond asymptotic expansions we compare
MCT-correlators for a finite distance from the critical point obtained as full
solutions to the equations (\ref{eq2}-\ref{eq4}). The fact that in
simulations only finite distances from the transition point are accessible
necessitates this comparison. We used  the monodisperse model
(B), because it provides reasonable nonergodicity parameters and contains
the barrier like the simulation (D).
\begin{centering}
\begin{figure}[h]
\centerline{\psfig{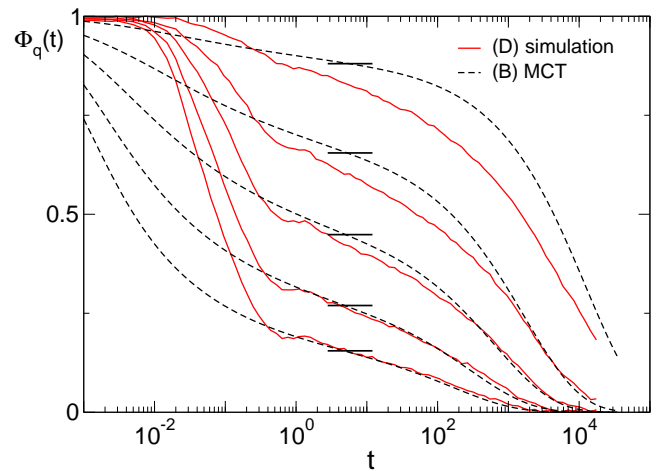}} \caption{
Density correlation functions in simulation ((D) red solid lines)
and MCT calculations in model (B) (black dashed lines): The
horizontal black bars indicate the critical plateau values of
$f_q^c$ in MCT. The wavevectors from bottom to top are $qD=57.1,
45.9, 33.9, 22.7, 12.5 $. The short time diffusion coefficient was
set to $D^s/D^2=0.133$. The theoretical curves correspond to a
polymer concentration of $\phi_p=0.2461$, which means a separation
parameter of about $\epsilon_{MCT}=-0.001$. The simulation results
were obtained for a polymer concentration $\phi_p=0.42$, which
corresponds to a separation parameter of $\epsilon_{simu}=-0.015$. }
\label{fig9}
\end{figure}
\end{centering}
The wave vectors for the
MCT-correlators have been chosen to be as close as possible to the
simulation values ($\Delta qD \pm 0.1$). The separation parameter for the
MCT calculation was adjusted so that the structural relaxation can be compared most succinctly.

 The correlators from simulation in
Fig. \ref{fig9} show a two-step relaxation process with the final
$\alpha$-relaxation from the plateau of height $f_q^c$. The
$\beta$-process describes the dynamics close to $f^c_q$ including
some part of the approach to $f^c_q$ from above. The simulation data
exhibit damped vibrational motion on short time scales. This is
neglected in \gl{eq2}, which therefore can describe the dynamics
only at later times. It is only this structural relaxation that MCT
addresses and thus the modelling at short times is done as simply as
possible.  We do not attempt to (a) include vibrational motion, (b)
capture the separation of short time and long time dynamics
quantitatively, but (c) only consider the shape of the structural
relaxation  in the following comparison.

The results highlight the strong q-dependence of the structural
relaxation. In addition a strong stretching, i.e. non-exponential
$\alpha$-relaxation is also observed. According to different
von-Schweidler exponents of $b=0.37$ in the simulation and $b=0.54$
in MCT the simulation results appear somewhat more stretched.
Nevertheless, the local dynamics, where the bond formation can be
directly seen, is well described by MCT. Amplitude and shape of the
$\alpha$-process are rather well captured, as would become even
clearer if simulations closer to the transition could be performed.
But for the dynamics on larger scales only qualitative statements
can be made .

\section{Conclusions}

Glass transitions within MCT are bifurcation points in the equations
of motion of the structural relaxation. While the equilibrium
structure of the considered glass forming fluid changes smoothly,
the dynamics slows down significantly and a metastable glassy
structure comes into existence. The bifurcation transitions of MCT
contain universal signatures, like von Schweider's law that is the
origin of the non-exponentiality of the (final or $\alpha$-)
structural relaxation. Non-universal amplitudes, like the critical
glass form factor $f^c_q$ or the critical amplitude $h_q$ entering
von Schweidler's law, contain the information in MCT about the
physical mechanisms causing arrest.

We considered the wavector dependence of $f^c_q$, $h_q$, and of the
$\alpha$ relaxation times $\tau_q$ in order to discover the origin
of glassy arrest in colloidal dispersions of particles with short
ranged attractions at intermediate packing fractions. The study was
motivated to gain insight into the connection between attraction
driven glass transitions at higher densities and colloidal gelation
at lower ones. We looked at simulations of a model system where
particles interact additionally with a weak long-range repulsive
barrier.

Structure factors $S_q$ directly taken from simulations were used as the
only input to MCT, in order to study the importance of the mesoscale
gel-like structure indicated by a pre-peak in $S_q$. It arises in
the simulations from the long-ranged repulsive barrier that
suppresses gas-liquid demixing. This mesoscale peak is the slowest
mode in the system, as the void-structure takes the longest time to
dissolve.  Yet, we find that the existence of the pre-peak in $S_q$
does not affect MCT calculations for the attraction driven glass
transition. The latter is caused by local bond formation apparent in
all quantities at large wavevectors. We conclude that the mesoscopic
mode is enslaved to the formation of  physical bonds, and that this
local process is not affected by the larger-scale heterogeneities of
the system. MCT quantiatively captures local bond formation but
somewhat overestimates the stability of the glass on larger length
scales. Still, the mechanism of arrest in the dispersion at
intermediate density is the attraction driven one discovered in MCT
at higher densities.

The role of polydispersity in MCT calculations was also  considered.
Averaged structure factors of polydisperse systems miss the large
$q$-tail indicative of short ranged attractions. This arises from
negative interference of the various contributions from
particle-pairs with different contact distances. While this prevents
the use of averaged $S_q$ from polydisperse systems to capture an
attraction driven glass transition, it does not imply that the
vitrification mechanism in polydisperse systems is different. Rather
the actual polydisperse system exhibits bond formation, and can be
described quantitatively within MCT using  the appropriate $S_q$
from, for example, the corresponding  monodisperse  system.

\begin{acknowledgments}

We thank M. Cates for valuable discussions. This work was supported by 'Acciones Integradas Hispano-Alemanas' of
the DAAD, by the  IFPRI-initiative  'Gelling Systems', and by the
DFG grants SFB~513, International Training Group 667, and  SP/714-3.
AMP acknowledges financial support from the MEC under project
MAT2006-13646-C03-02, and JB acknowledges financial support by the
IUF. 

\end{acknowledgments}

\end{document}